\documentclass[a4paper]{article}

\usepackage{graphicx}
\usepackage{bm}

\begin{document}

\title{Memristive switching of MgO based magnetic tunnel junctions}
\author{Patryk Krzysteczko$^*$, G\"unter Reiss, Andy Thomas\\Bielefeld University, Thin Films and Physics of Nanostructures,\\33615 Bielefeld, Germany}

\date{}

\maketitle





\textbf{The search for nonvolatile memory concepts has a massive impact on the development of nanoscopic systems with adjustable electrical properties. Capacitor-like structures composed of insulating materials sandwiched between metallic electrodes are envisioned to overcome the limitations associated with conventional charge storage devices and may open the road to neuromorphic computing. Together with phase-change memories \cite{ovshinsky1986prl,wuttig2007nat} two concepts attracted extensive interest \cite{meijer2008science}: resistive and magnetoresistive random access memories (RAM). In resistive RAM, electrochemical processes within the insulating layer enable the access to multiple resistive states \cite{liu2000apl, beck2000apl, szot2006nat, waser2007nat}. For magnetic RAM, ferromagnetic electrodes are used to switch the resistance by changing the alignment of the electrodes from parallel to anti-parallel and vice versa \cite{PRL1995V74S3273,prb2001V63S054416,chappert2007nat}. Here we demonstrate that both effects can be observed simultaneously in nano-scale magnetic tunnel junctions (MTJs). The framework of a 2nd order memristive system \cite{chua1976ieee} is utilized to treat the tuning of electrical resistance by concurrent modulation of two material parameters.}








\section*{ }

The development of nanoelectronic nonvolatile memories and neuromorphic computer architectures requires devices with stable resistance states and analogue capabilities to mimic, e.g., neurobiological architectures. Both concepts have large significance for development and improvement of functionalized materials and nano-scaled systems with adjustable electric properties. Specifically engineered memristors provide controllable resistance where the definition of the memristor is based solely on fundamental circuit variables. The Memristor theory was formulated by Leon Chua in 1971 \cite{chua1971ieee}. Chua extrapolated the conceptual symmetry between the resistor, inductor, and capacitor, and inferred that the memristor is a similarly fundamental device. Other scientists had already used fixed nonlinear flux-charge relationships, but Chua's theory introduces generality. The interest in this concept renewed in 2008 due to experimental progress by Strukov et al.\ \cite{strukov2008nat} where a memristor was realized as a Pt/TiO/Pt sandwich. Moreover, Strukov et al. created a link between the pioneering work of Chua and the field of resistive switching. The memristor ``remembers'' the history of the applied voltage $v$ and current $i$ since its resistivity depends on the time integrals of $v$ and $i$. A voltage controlled memristor can be described by
\begin{equation}\label{memristor}
u(t)=R\left(\Phi(t)\right)\cdot i(t)
\end{equation}
where $\Phi(t)=\int_0^t u(t')dt'$ is referred to as flux and $R$ ist the (memory) resistance \cite{yang2008nat,strukov2009small}. 
The memristor was generalised to memristive systems in a 1976 publication by Chua and Kang \cite{chua1976ieee}. Whereas a memristor has mathematically scalar state, a memristive system has vector state. The number of state variables is independent of, and usually greater than, the number of terminals. In this paper, Chua applied this model to empirically observed phenomena, including the Hodgkin-Huxley model of the axon and a thermistor at constant ambient temperature. He also described memristive systems in terms of energy storage and easily observed electrical characteristics. These characteristics match resistive random-access memory and phase-change memory, relating the theory to active areas of research. In the more general concept of an \textit{n}th order memristive system, equation\,\ref{memristor} turns to
\begin{eqnarray}\label{memristive}
u(t)&=&R\left(\bm{w},u,t\right)\cdot i(t)\\
\dot{\bm{w}}&=&f(\bm{w},u,t) 
\end{eqnarray}
where $\bm{w}$ is a set of $n$ state variables describing the device \cite{chua1976ieee,yang2008nat,strukov2009small}. Neither $R$ nor $f$ are explicit functions of $u$ in the case of a ``pure'' memristor.

\begin{figure}
	\begin{center}
 	\includegraphics{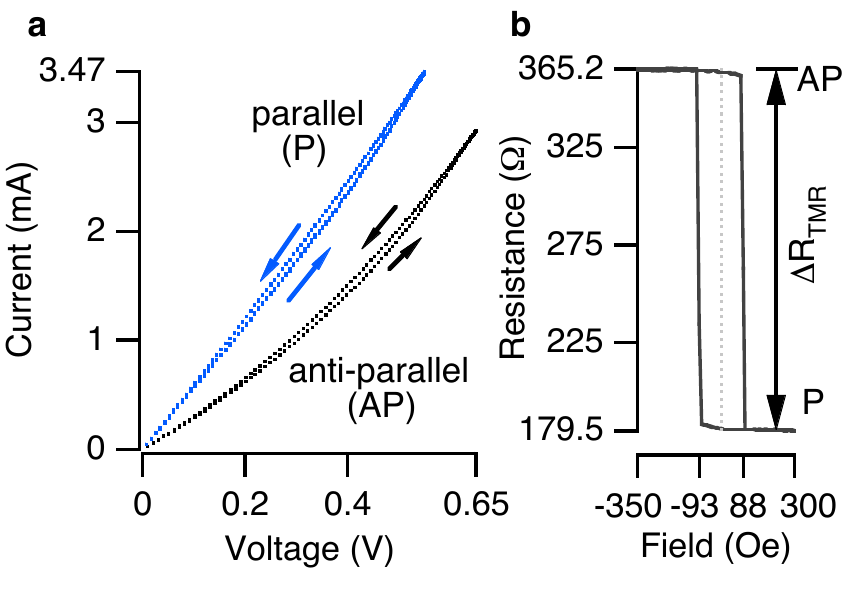} 
	\end{center}
	\vspace{-0.75cm}
	\caption{\textbf{Simultaneous occurrence of resistive and magnetoresistive switching.} \textbf{a}, A slight splitting of the $i$-$v$ curve can be observed for both magnetic states, which demonstrates the presence of resistive switching. Since the curves are highly symmetric with respect to the origin, only the first quadrant is shown. \textbf{b}, The magnetoresistive switching of the device characterized by a magnetic minor loop. (see also Supplementary Information Fig.\,1 and 2)} 
	\label{iv}
\end{figure}

After substantial effort was invested in studies of resistive and magnetoresistive systems, the next step is a combination of both effects in one device \cite{ventura2007jpd,halley2008apl,yoshida2008apl}. These systems can be treated as 2nd order memristive systems with $w_1$ the magnetic state of the electrodes and $w_2$ describing the state of the barrier. In this framework, we present studies on MgO based low resistive magnetic tunnel junctions with Co$_{66}$Fe$_{22}$B$_{12}$ electrodes. The samples are characterized by tunnel magneto resistance (TMR) ratios of about 100\,\% and exhibit an additional bipolar resistive switching (RS) of up to 6\,\%. The TMR ratio is not reduced by the operation as resistive memory and the area resistance is low ($6.8\,\Omega\mu m^2$). Therefore, the presented results are a leap towards a combined magnetoresistive and resistive memory that could be switched by spin transfer torque \cite{wang2009ieee}. Figure~1 shows the current-voltage ($i$-$v$) curves for the parallel (P) and anti-parallel (AP) orientation of the Co-Fe-B electrodes. The current value measured for increasing voltage is slightly lower then the current measured for decreasing voltage. This hysteresis of the $i$-$v$ curve is observed in both magnetic sates and is the primary characteristic of memristive behaviour.

\begin{figure}
	\begin{center}
 	\includegraphics{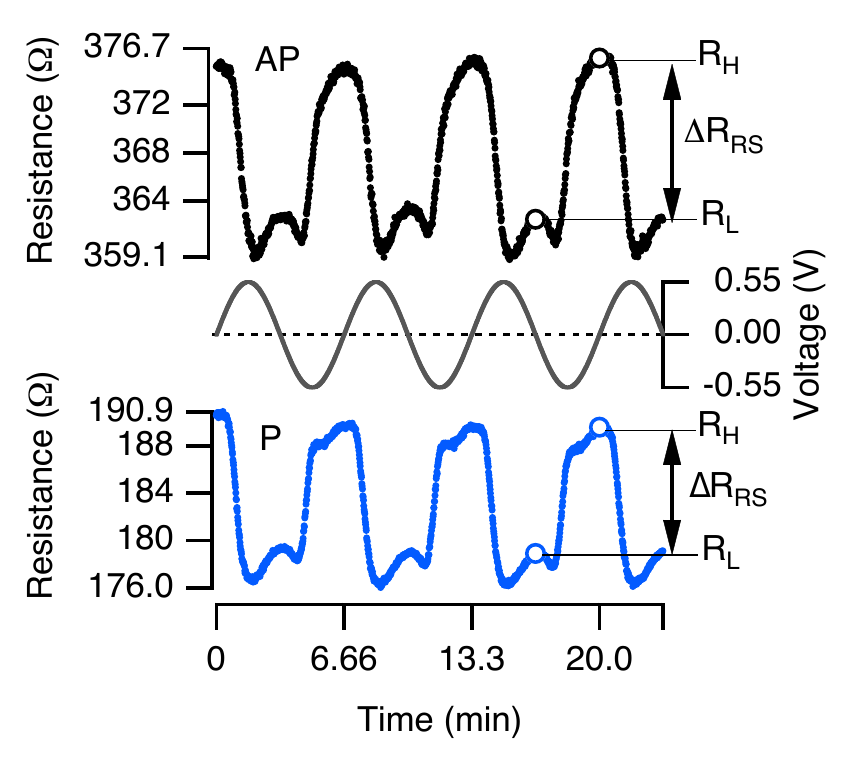}
	\end{center}
	\vspace{-0.75cm}
	\caption{\textbf{Resistive switching (RS) of a magnetic tunnel junction.} The RS is induced by the sinusoidal voltage plotted at the right axis. The resistance of both magnetic states (AP and P) is switched periodically between two resistive states ($R_{\rm H}$ and $R_{\rm L}$). Labels are attached to the resistance curves (at $V=0$) to clarify the definition of $R_{\rm H}$ and $R_{\rm L}$. A magnetic field of $\rm\pm400\,Oe$ provides a stable parallel or anti-parallel orientation of the magnetic layers which suppresses spin transfer torque effects during the RS measurements. The read-out delays are neglected in the time scale.} 
	\label{rt}
\end{figure}

To further investigate this effect, we reduce the applied voltage once a second and measure the resistance at $20\rm\,mV$ with a read-out delay of $\rm 200\,ms$. This procedure removes non-linear contributions resulting from the voltage-dependent tunnel resistance \cite{brinkman1970jap} and is typical for nonvolatile memories where the information is written at high and read at low bias. Figure~2 shows the variation of the MTJ's resistance induced by the application of sinusoidal voltage. The resistance of the AP state is switched between a high resistance ($R_{\rm H}$) and a low resistance ($R_{\rm L}$) state. A similar resistive switching is observed for the P state. The high and low resistance states are defined by the resistance measured at zero voltage, as indicated by the circles in Figure~2 and 3. Table~\ref{table} summarizes the results obtained from the data presented in Figure~2.

\begin{table}[!h]
\begin{tabular}{lcccc}
					 & $R_{\rm H}$ & $R_{\rm L}$ & $\Delta R_{\rm RS}$ & RS \\
\hline
$R_{\rm AP}$			& 375.9$\,\Omega$ & 362.5$\,\Omega$ & 13.4$\,\Omega$ & 3.7\% \\
$R_{\rm P}$			& 189.6$\,\Omega$  & 178.9$\,\Omega$  & 10.7$\,\Omega$ & 6.0\% \\
$\Delta R_{\rm TMR}$	& 186.3$\,\Omega$ & 183.6$\,\Omega$ &  & \\
TMR					 & 98.3\% & 102.6\% & & \\
\end{tabular}
\caption{Resistive (RS) and magnetoresistive (TMR) effects calculated from the data in Fig.~2. The index to $R$ is denoting the (magneto)resistive state of the device. $\Delta R$ is the total resistance change due to the effect denoted by the index.}\label{table}
\end{table}

\begin{figure*}
	\hspace{-2cm}
	\includegraphics{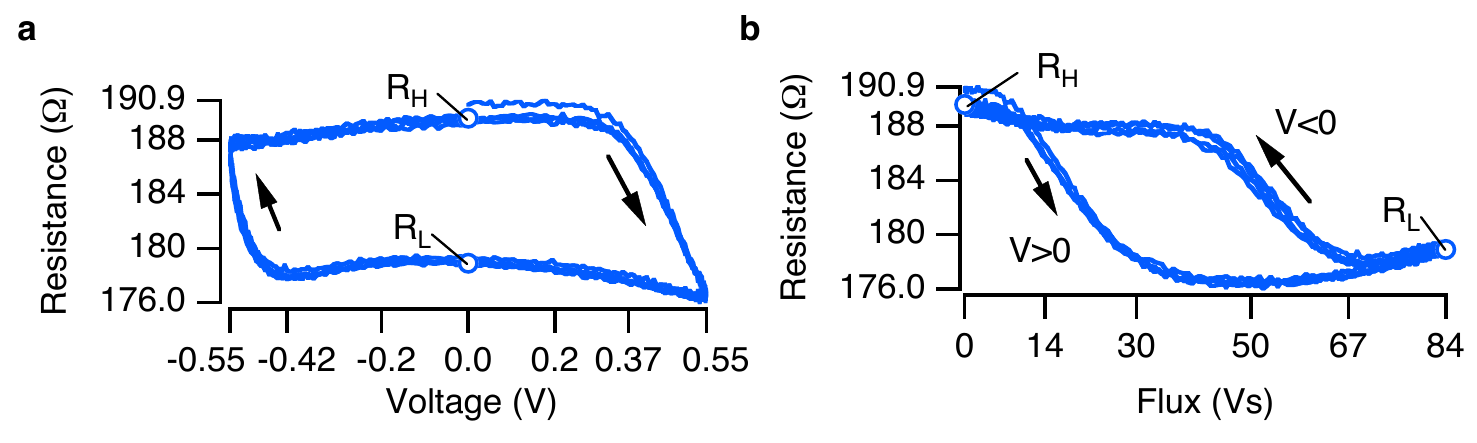}
	\vspace{-0.75cm}
	\caption{\textbf{Two complementary methods to display the resistance of the P state.} 2.5 resistance loops are plotted on top of each other. \textbf{a}, The bipolar resistive switching between two stable resistance states becomes clear when plotted vs.\ voltage. \textbf{b}, The hysteretic nature of resistive switching is is shown with clarity when plotted vs.\ flux.} 
	\label{flux}
\end{figure*}

Figure~3a shows the resistance of the P state now as a function of the applied voltage. Starting at $R_{\rm H}$ the resistance remains almost constant until a critical voltage of about $0.37\rm\,V$ is reached. Then, the resistance drops reaching the lowest value for maximum bias voltage. A slight increase of the resistance can be observed when the applied voltage is released. A similar switching process appears in opposite direction after the polarity of the applied voltage is changed and, again, when a critical voltage for the switching is reached ($-0.42$\rm\,V). Figure~3b shows the hysteretic nature of the resistance as a function of flux. The flux ``applied" to the sample increases when a positive voltage is applied and decreases for negative polarity. For increasing flux the resistance decreases slightly till a flux of about $14\rm\,Vs$ is reached and the resistive switching begins. The minimum resistance is reached at about $42\rm\,Vs$ before the resistance increases slowly reaching $R_{\rm L}$ at $84\rm\,Vs$. The back-switching begins at $67\rm\,Vs$ leading to a pronounced resistive hysteresis. The resistive loops lie on top of each other, indicating the reproducible nature of the switching process.


\begin{figure}[!h]
	\begin{center}
 	\includegraphics{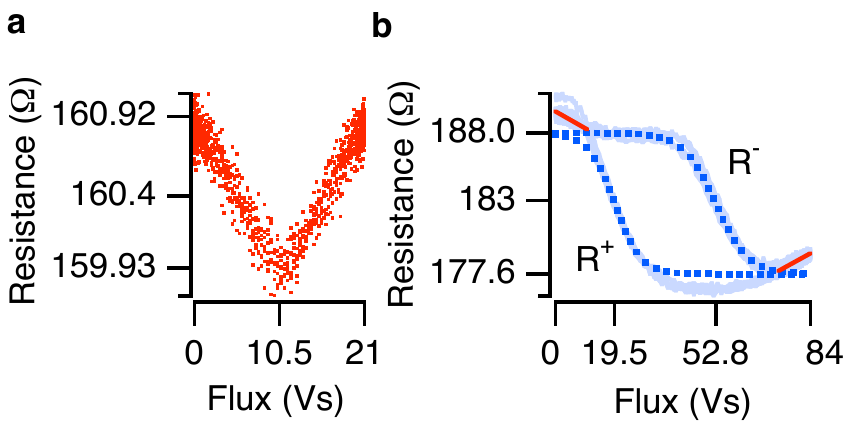}
	\end{center}
	\vspace{-0.75cm}
	\caption{\textbf{Thermal contribution to the $\bm{R(\Phi)}$ curve.} \textbf{a}, Reference measurement on an MTJ showing no RS. The amplitude of the applied voltage was 350\,mV, 2.5 periods are plotted on top of each other. \textbf{b}, The fitting-curves calculated by Eq.\,\ref{fit} are shown by dotted lines while the experimental data, the same as in Fig.~3b, is given by the curve in the background. The deviation of the experimental data from the fitting-curve due to thermal contributions is depicted by red lines.} 
	\label{thermo}
\end{figure}

As pointed out previously \cite{krzysteczko2009jmmm}, a possible explanation for the RS of MTJs is a displacement of oxygen vacancies located at the bottom Co-Fe-B/MgO interface. We perform measurements on MTJs with an MgO barrier prepared by post-oxidation of an Mg layer to study this model in more detail. The bottom Co-Fe-B electrode is not exposed to oxygen for this preparation method and no oxygen vacancies at the bottom interface are expected \cite{mayerheim2002prb}. The corresponding measurement is shown in Figure~4a. Two features are noted. First, no $R(\Phi)$-hysteresis can be observed indicating the absence of resistive switching, as expected. Second, a V-shaped curve is observed with the resistance lowered by 0.62\,\% in the middle of the flux-axis. At this flux the applied voltage reaches its maximum for both polarities. Since the tunnelling elements (and the top and bottom conduction lines) are heated due to the high current density the local temperature may not relax during the 200 ms of read-out delay. Therefore, a possible explanation for the reduced resistance is an elevated temperature of the MTJ. A similar influence of remnant heat is visible for the sample showing a distinct RS behaviour as denoted by the red lines in Figure~4b.

\begin{figure}
	\begin{center}
 	\includegraphics{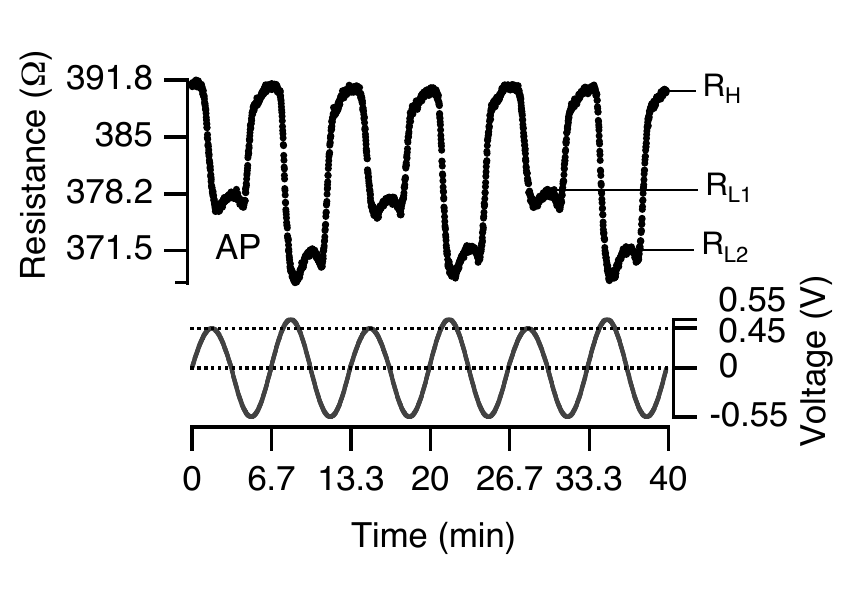}
	\end{center}
	\vspace{-0.75cm}
	\caption{\textbf{Influence of an alternating voltage amplitude on the $\bm{R_{\rm L}}$ state}. The positive amplitude is periodically switched between $450\rm\,mV$ and $550\rm\,mV$. This leads to a splitting of the low resistance state into $R_{\rm L1}$ and $R_{\rm L2}$ separated by $6.7\,\Omega$.} 
	\label{two_level}
\end{figure}

Now, we stress the question how the presented system differs from a pure memristor. Despite the superposition with thermally induced resistance changes, the hysteresis of $R(\Phi)$ means that a pure memristor can be defined at best ``branch-wise'':
\begin{equation}\label{fit}
R^\pm(\Phi)=R_{\rm H}-\frac{\Delta R_{\rm RS}}{1+\exp\left((\Phi_{\rm S}^\pm-\Phi)/\delta^\pm \right)}
\end{equation}
with $R^+(\Phi)$ the resistance on the positive branch ($V>0$), $\Phi^+_{\rm S}=19.5\rm\,Vs$ the inflection point and $\delta^+=4.3\rm\,Vs$ indicating the sharpness of the switching process. To fit the negative branch $R^-(\Phi)$ we shift the inflection point to $\Phi^-_{\rm S}=52.8\rm\,Vs$ and use $\delta^-=5.3\rm\,Vs$. Both fitting curves are shown in Figure~4b. However, a simple hysteresis as presented in Figure~3 will only result from a sinusoidal voltage of constant amplitude. If the positive amplitude is alternated between two values, as illustrated in Figure~5, an additional $R_{\rm L}$ state is generated. In general, multiple resistance states both high and low can be created by according amplitude variation.


In summary, simultaneous magnetoresistive and bipolar resistive switching of MTJs is presented. The entire structure can be understood in the framework of an 2nd order memristive system. Five resistance states are demonstrated for one MTJ proving that the combination of resistive and magnetoresistive switching of MTJs provides a method for multi-bit data-storage and logic.

\section*{Methods}

The TMR systems are sputter deposited in a Singulus \textsc{timaris II} tool. The film sequence is (bottom) Ta 3/Cu-N 90/Ta 5/Pt$_{37}$Mn$_{63}$ 20/Co$_{70}$Fe$_{30}$ 2/Ru 0.75/Co$_{66}$Fe$_{22}$B$_{12}$ 2/MgO 1.3/Co$_{66}$Fe$_{22}$B$_{12}$ 3/Ta 10/Cu 30/Ru 7 (top). The thickness is given in nm and the composition in at. \%. Elliptical TMR elements with sizes in the range of $\rm 0.038\pm0.005\,\mu m^2$ are prepared by electron beam lithography in combination with ion beam etching (see Supplementary Information Fig.\,3). The film sequence for the reference structure showing no RS behaviour is (bottom) Ta 5/Cu-N 90/Ta 5/Pt$_{37}$Mn$_{63}$ 20/Co$_{70}$Fe$_{30}$ 2.2/Ru 0.8/Co$_{60}$Fe$_{20}$B$_{20}$ 2/MgO 0.8/native ox. 1\,torr 1200\,s/Mg 0.3/Co$_{60}$Fe$_{20}$B$_{20}$ 1.5/Ta 10/Cu 30/Ru 7 (top).

All transport measurements are performed using a constant voltage source at room temperature with the bottom electrode grounded. To induce RS we apply voltage pulses of up to $v_{\rm max}=0.65\,\rm V$ for one second to stress the samples by a current density of typically $8\times 10^6\rm\,A/cm^2$. All resistance values are measured at a read-out voltage of $20\rm\,mV$ with a read-out delay of $\rm 200\,ms$ after each stress pulse. In total, the applied voltage sequence is $v_n(t)=v_{\rm max}\sin (t)$ for every odd pulse-number $n$ and $v_n=20\,\rm mV$ for even pulse-numbers (see Supplementary Information Fig.\,4).


\section*{Acknowledgements}

The Authors gratefully acknowledge Singulus Technologies AG for providing the samples.

\newpage

\section*{Supplementary information}

\begin{figure}[!h]
 	\includegraphics[width=1\linewidth]{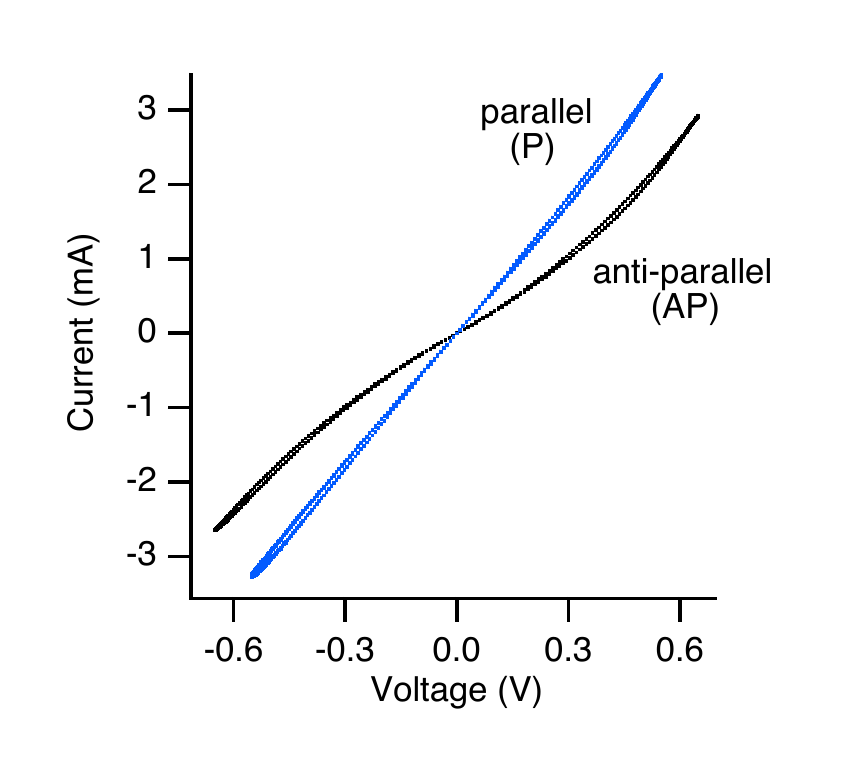}
	
	\vspace{0cm}\caption{A slight splitting of the $i$-$v$ curve can be observed for both magnetic states which demonstrates the presence of resistive switching. Since the curves are highly symmetric with respect to the origin, only the first quadrant is shown in the main article.} 
	\label{iv}
\end{figure}
%

\begin{figure}
 	\includegraphics[width=1\linewidth]{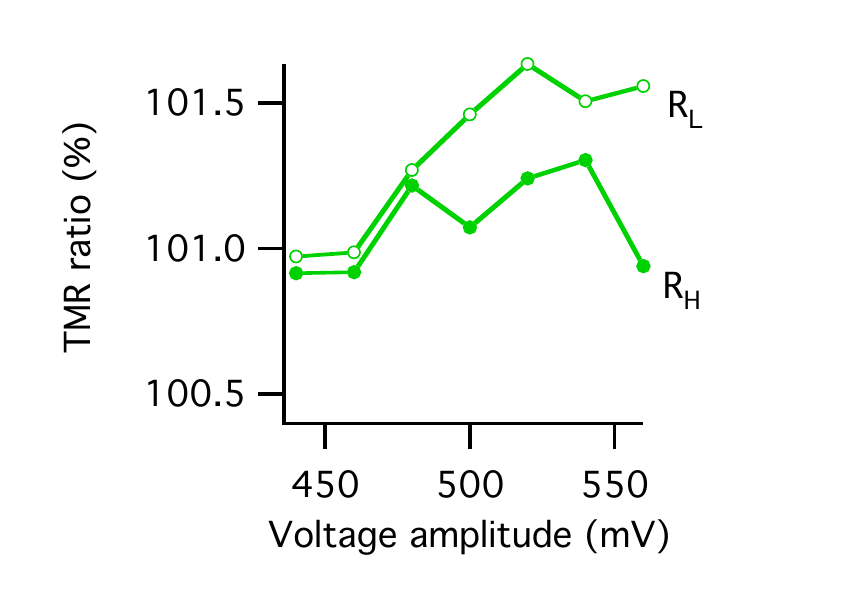}
	
	\vspace{0cm}\caption{Resistive switching without barrier degradation. To get the presented data the MTJ is switched 5 times between the high resistive state $R_{\rm H}$ and the low resistive state $R_{\rm L}$. The TMR ratio is obtained from an magnetic minor loop measured in the $R_{\rm L}$ state (open circles). Again, the MTJ is switched 5 times and the TMR ratio is measured, now in the $R_{\rm H}$ sate (solid circles). Then the voltage amplitude is increased by $250\,\rm mV$ and the procedure repeated. Since the TMR ratio remains high, a degradation of the MgO barrier due to resistive switching can be excluded.} 
	\label{iv}
\end{figure}
%

\begin{figure}
 	\includegraphics[width=1\linewidth]{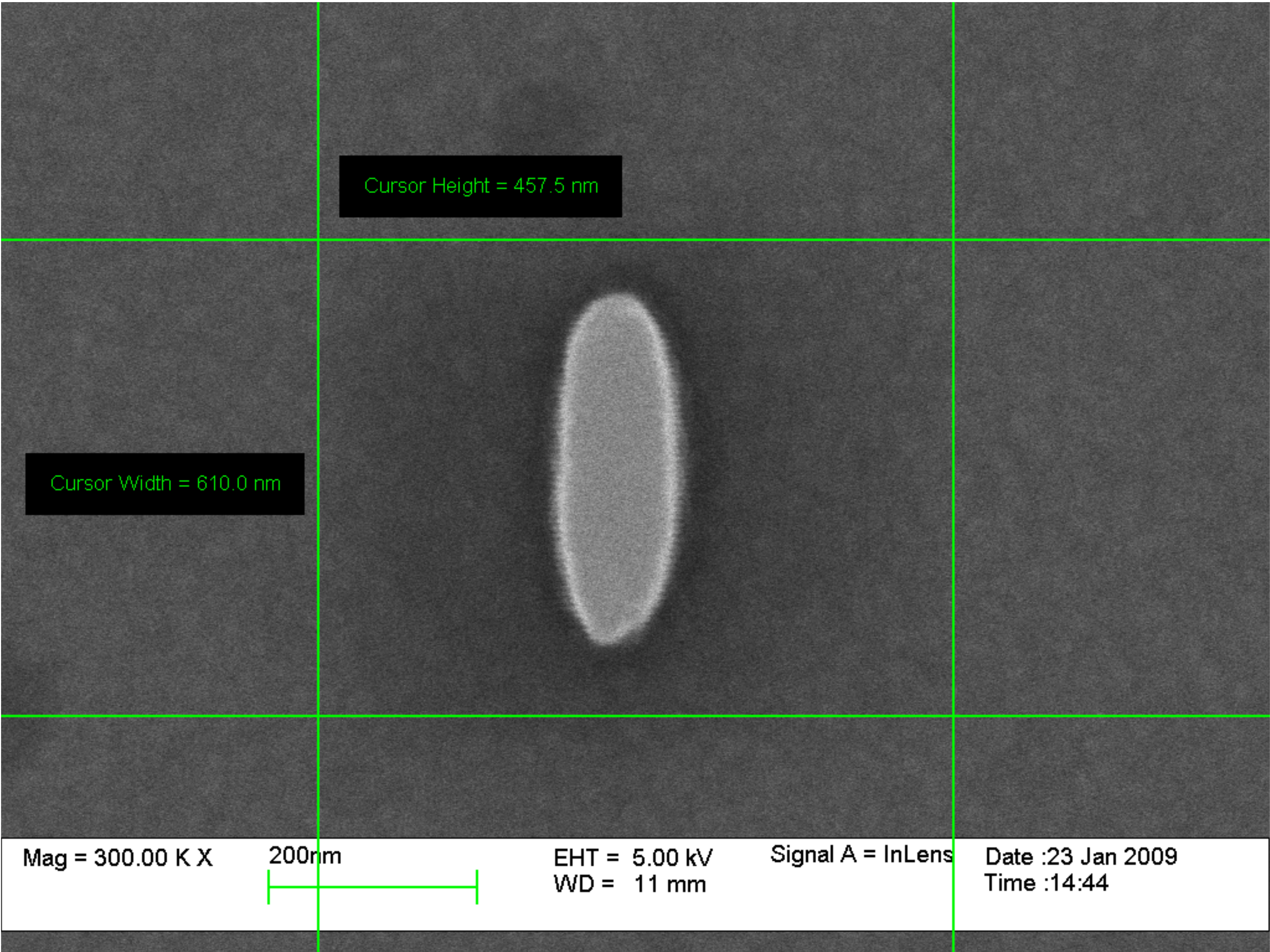}
	\includegraphics[width=1\linewidth]{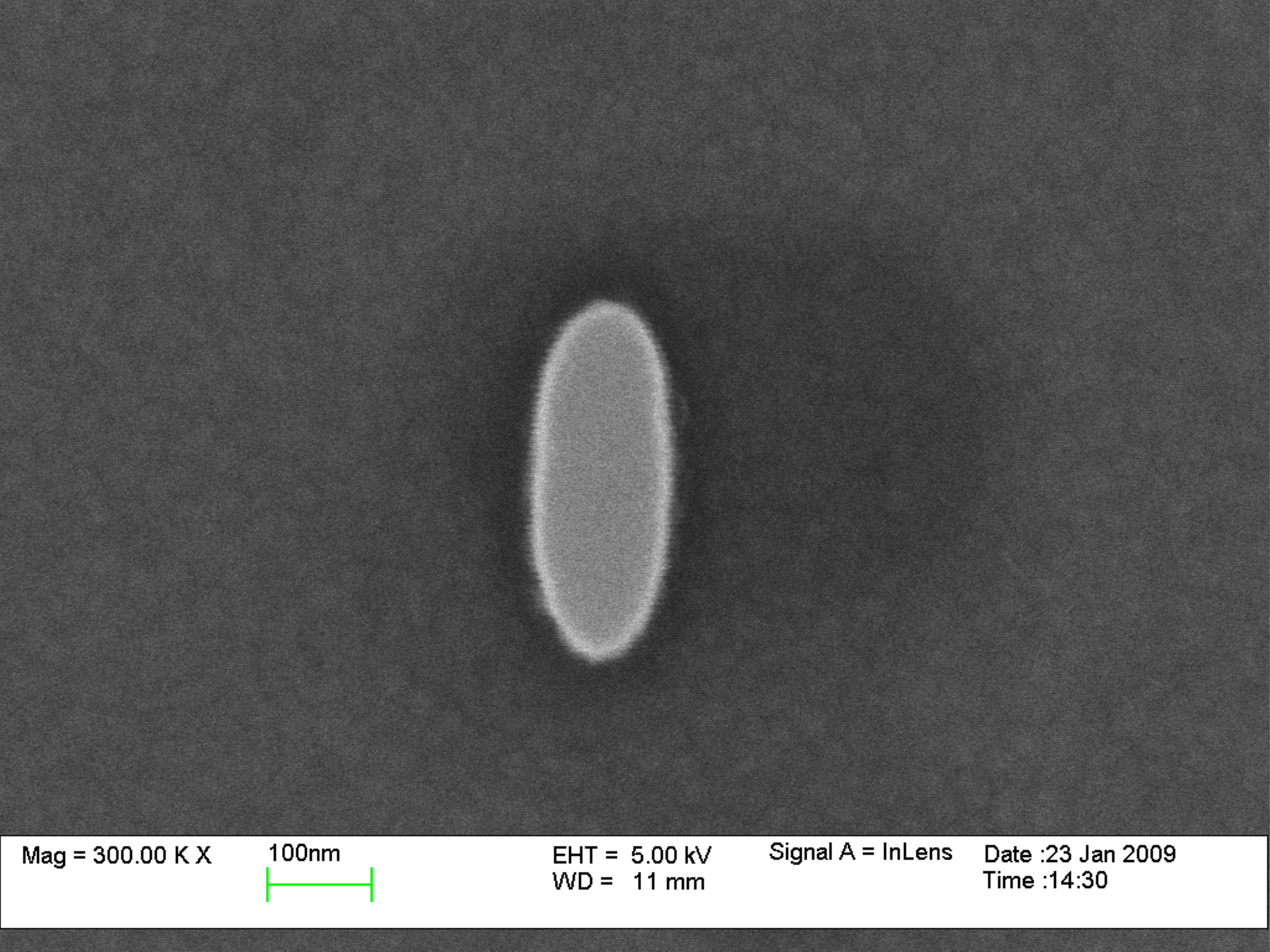}
	\caption{Scanning electron microscope (SEM) micrographs of typical elements prior to etching. The width of the resist mask is 119\,nm $\times$ 337\,nm (top) and 130\,nm $\times$ 343\,nm (bottom).} 
	\label{iv}
\end{figure}
%

\begin{figure}
 	\includegraphics[width=1\linewidth]{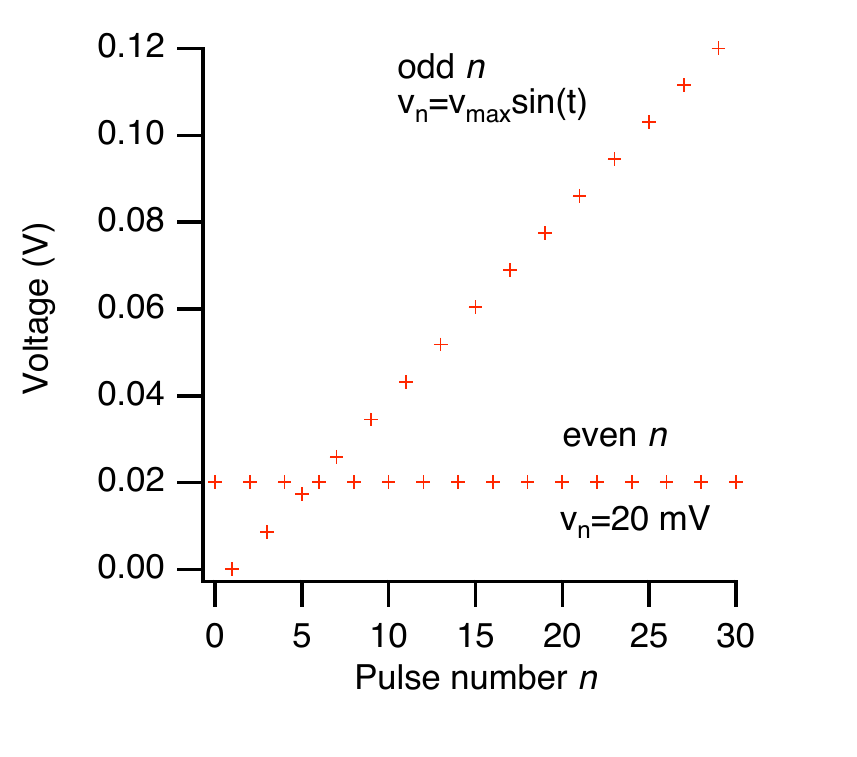}
	
	\vspace{0cm}\caption{Details of the measurement procedure. The applied voltage sequence is $v_n(t)=v_{\rm max}\sin (t)$ for odd pulse-numbers $n$ and $v_n=20\,\rm mV$ for even pulse-numbers. The length of the odd pulses is 1000\,ms, the length of the even pulses is 200\,ms. All resistance values are obtained after an odd (delay) pulse at 20\,mV,  whereas the resistive switching is induced by the even voltage pulses.} 
	\label{iv}
\end{figure}

\newpage

\end{document}